\documentclass[12pt,a4paper,twoside]{article}
\usepackage{epsfig}
\usepackage{baltlat1}
\usepackage{wrapfig}
\begin{document}
\ \
\vspace{0.5mm}

\setcounter{page}{1}
\vspace{8mm}

\titlehead{Baltic Astronomy, vol.12, XXX--XXX, 2003.}

\titleb{ESO\,422-G028: THE HOST GALAXY OF A GRG}

\begin{authorl}
\authorb{M.~Jamrozy}{1}
\authorb{J.~Kerp}{1} 
\authorb{U.~Klein}{1} 
\authorb{K.-H.~Mack}{2, 1} and
\authorb{L.~Saripalli}{3} 
\end{authorl}

\begin{addressl}
\addressb{1}{Radioastronomisches Institut der Universit{\"a}t 
Bonn, Auf dem H{\"u}gel 71, D-53121 Bonn, Germany }

\addressb{2}{Istituto di Radioastronomia, Via P. Gobetti 101, 
I-40129 Bologna, Italy}

\addressb{3}{Australia Telescope National Facility, Locked Bag 194, 
Narrabri NSW 2390, Australia}
\end{addressl}

\submitb{Received: xx October 2003}

\begin{abstract}
 ESO\,422-G028 denotes the very center cD-galaxy of the giant radio 
source B0503-286 (Saripalli  et al. 1986, Subrahmanya \& Hunstead, 
1986). The angular extent of the associated radio structure is about 42\farcm4, 
which corresponds to a linear size of 1.89~Mpc 
(with {\footnotesize$\rm\Omega_{M}=0.27$}, 
{\footnotesize$\Omega_{\Lambda}=0.73$}, and 
{\footnotesize$H_{0}=71$}~km~s{\footnotesize$^{-1}$}~Mpc{\footnotesize$^{-1}$}). 
Here we present new high-frequency total-power 
and polarization radio maps obtained with the Effelsberg 100-m dish. 
In addition, we correlate the radio data with optical and X-ray 
observations to investigate the physical conditions of both, the host 
galaxy and the extended structure.
\end{abstract}


\resthead{ESO\,422-G028:  The Host Galaxy of a GRG}
{M.~Jamrozy, J.~Kerp, U.~Klein,  K.-H.~Mack, L.~Saripalli}

%{Institution}{Author(s)}

%\def\ninepoint{\def\rm{\fam0\ninerm} \textfont0=\ninerm}

\sectionb{1}{HOST GALAXY}
The giant radio galaxy (GRG) B0503-286 is hosted by a massive galaxy
ESO\,422-G028 which has a redshift of 0.0381. The galaxy has a mean size 
of about 60~kpc and is the largest member of a poor galaxy cluster. Its optical 
apparent B and R magnitudes (taken from the Lyon-Meudon Extragalactic Data Base) 
are 14\fm62 and 13\fm01, respectively. 
Trifalenkov (1994) found that the galaxy contains a huge 
({\footnotesize$10^{6} - 10^{7}M_{\odot}$}) amount of warm (35--50~K) dust. Veron-Cetty \& Veron 
(2000) found that the nucleus of ESO\,422-G028 has a low luminosity and 
hence classified this galaxy as a LINER type. A low-resolution optical 
spectrum taken by Subrahmanya \& Hunstead (1986) and by Saripalli et al. 
(1986) does not show any H{\footnotesize$\beta$} and [OIII] emission lines, both are typical 
for powerful radio galaxies.
                
%\newpage
\sectionb{2}{HIGH FREQUENCY RADIO EMISSION}
The radio source B0503-286 consists of a weak  core-jet structure 
({\footnotesize$S_{1.4~GHz}$} {\footnotesize$\sim 12$~mJy}), well visible 
in the NVSS map (Condon et al.  1998), as well as of two prominent radio lobes 
elongated north-south 
({\footnotesize$S_{1.4~GHz}$} {\footnotesize$\sim 2.5$~Jy}). Morphologically it 
is a standard FRII-type (Fanaroff \& Riley 1974), but because of its rather low 
total radio luminosity ({\footnotesize$\log L_{1.4\rm ~GHz}=25.2$} {\footnotesize W~Hz$^{-1}$}) 
it has to be classified as a FRI/FRII transient type (Owen \& Ledlow 1994).

\begin{figure}
\centerline{{\psfig{figure=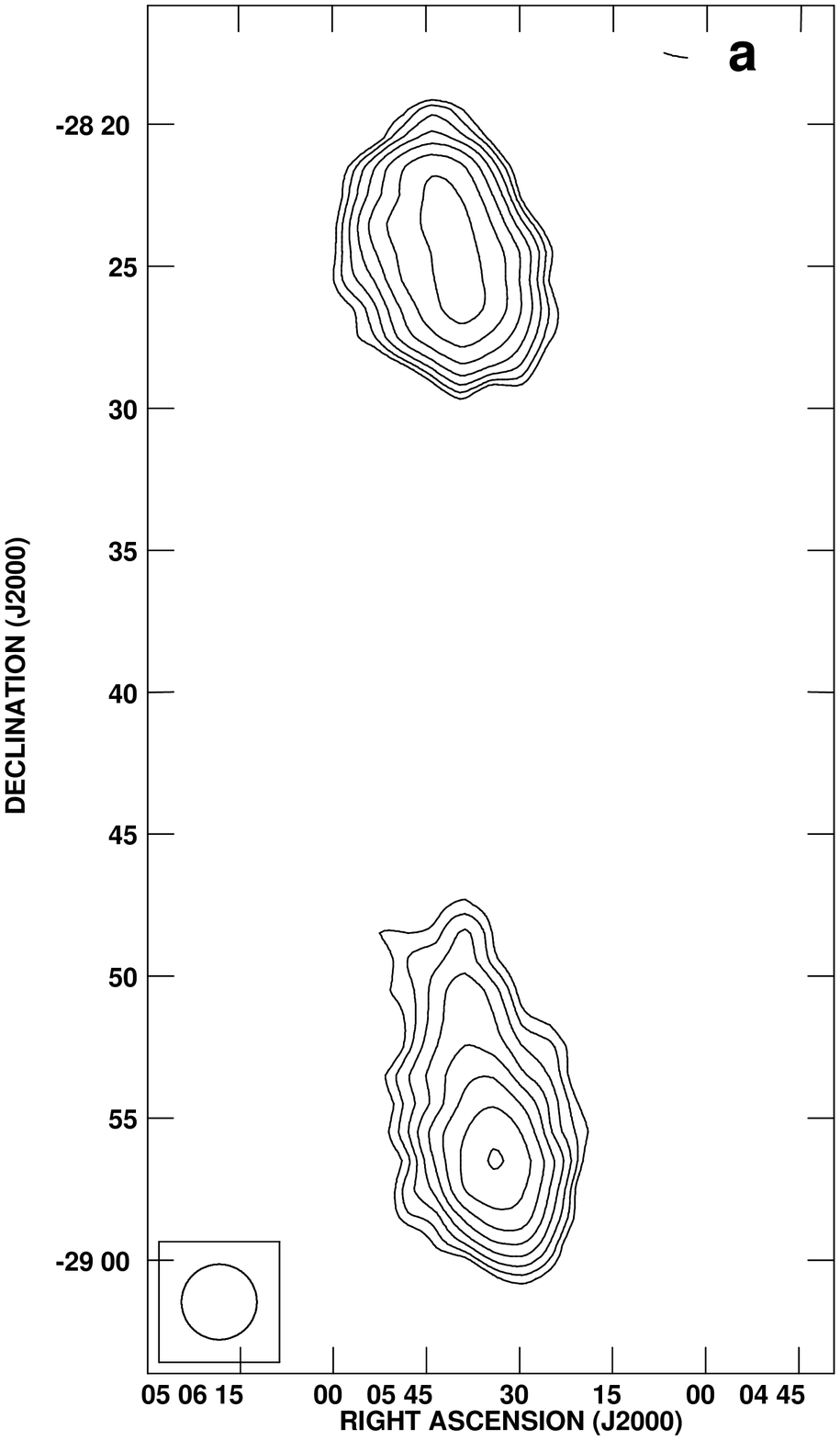,width=35truemm,angle=0,clip=}}
\hskip2mm
{\psfig{figure=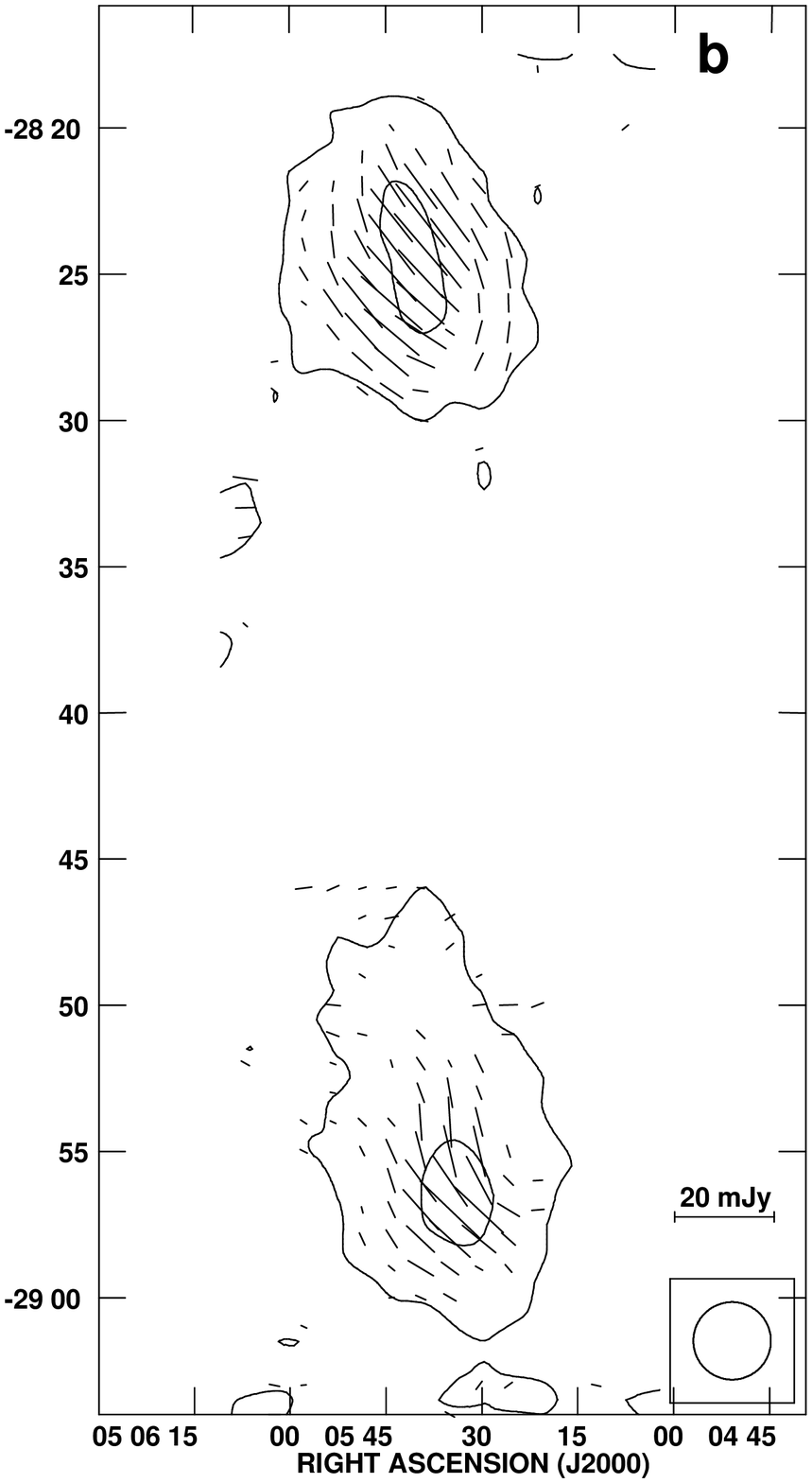,width=35truemm,angle=0,clip=}}
\hskip2mm
{\psfig{figure=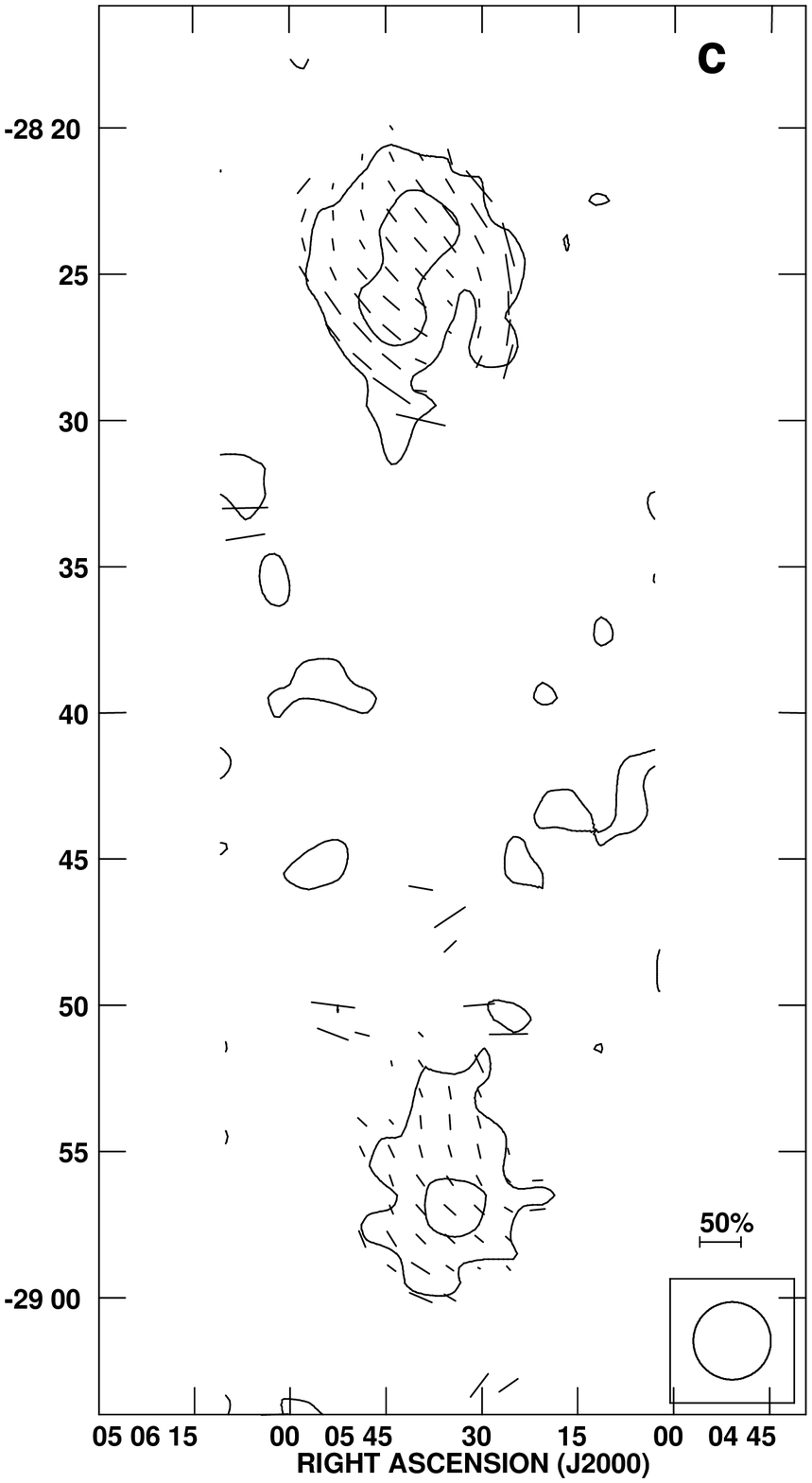,width=35truemm,angle=0,clip=}}}
\captionb{1}{Radio maps of the source B0503-286 at {\footnotesize$\lambda$}6.3~cm. 
{\bf a:} Total intensity contours of 1, $\sqrt{2}$, 2, ...,
$\sqrt{2}^{7}$  $\times$ 10~mJy/beam.
{\bf b:} Magnetic field B--vectors, with their length proportional 
to the polarized intensity, superimposed with 5 and 80~mJy/beam total intensity 
contours. {\bf c:} Linearly polarized intensity contours of 3 and 
12~mJy/beam, with the vectors of the fractional linear polarization superimposed. 
The scales of the polarization vectors are indicated by the bars and the beam size 
by the circles in the lower corners of the maps.}
\end{figure}

Here we report new high-frequency measurements of the radio giant, carried 
out with the Effelsberg 100-m telescope. The {\footnotesize$\lambda\lambda$} 
6.3 and 2.8~cm observations are essentially free of Faraday effects and hence 
allow us to directly map the intrinsic magnetic field structure and polarization 
characteristics of the source. In addition, the high-frequency maps (see 
Fig.~1) emphasize the most active parts of the radio galaxy which are very 
important when studying particle ageing. The linear polarization maps reveal 
a significant increase of the fractional polarization as well as a tangential 
magnetic field structure in the region of the hot spots. Furthermore, the 
lobes show high and almost equal fractional polarization, which is almost 
constant at both frequencies. Integrating the Stokes Q and U maps yields a 
fractional polarization of B0503-286, of $\sim$12 per cent, which is almost twice 
the median value that comes from the B3-VLA sample of radio sources (Klein 
et al. 2003). The ratio of the lobe flux densities is about unity at low as 
well as at high frequencies. The spectral index {\footnotesize$\alpha$} 
({\footnotesize$S\sim\nu^{-\alpha}$}) 
of the northern and southern lobe is 1.14 and 1.05, respectively. We do not 
detect any radio emission from the core of B0503-286 at 
{\footnotesize$\lambda$}6.3~cm. However, we manage to measure its flux density 
at {\footnotesize$\lambda$}2.8~cm. The resulting spectral index of the core 
between {\footnotesize$\lambda\lambda$} 21~cm and 2.8~cm is 0.34.

\begin{center}
\vbox{\footnotesize
{\smallbf\ \ Table 1.}{\small\ High-frequency radio flux-densities of 
the total source and its components.}
\begin{tabular}{rcccc}
 & & & & \\
\hline
\multicolumn{1}{c} {{\footnotesize$\lambda$}}   & \multicolumn{4}{c} {Flux} \\
\multicolumn{1}{c} {[cm]}  & \multicolumn{4}{c} {[Jy]} \\
                            &Total&N--lobe&S--lobe&Core \\
\hline
6.3 & $0.681\pm0.038$  & $0.346\pm0.028$ & $0.335\pm0.026$ &    -             \\
2.8 & $0.369\pm0.011$ &$0.193\pm0.008$  & $0.170\pm0.008$ &$ 0.006\pm0.002$ \\
\hline
\end{tabular}
}
\end{center}

\sectionb{3}{X-RAY EMISSION AND INTERGALACTIC MEDIUM}
We retrieved {\it ROSAT} PSPC observations toward the source\\ B0503-286
from the public {\it ROSAT} archive. The total exposure time of the 
observation is 21,825~s. The photon events were binned into the standard 
C, M and J {\it ROSAT} energy bands (for details see Kerp et al. 2002). 
We classified an X-ray source as detected if the number of net counts exceeded 
the 3$\sigma$ threshold above the noise set by the background level.
We integrated the X-ray photons of each individual source within a 
circular area with a diameter equal to 1$\farcm$0 for each individual 
energy band. From this total number of photons we subtracted the 
contribution of the X-ray background (XRB) emission.  In total, we detected 
27 significant X-ray sources within the boundaries of the radio giant. 
Their positions are marked in Fig.~2. as circles. The circles with a cross mark 
X-ray sources which lack 
optical counterparts brighter than {\footnotesize$R=19^{m}$} and which might be 
physically connected with the radio structure of the northern lobe. In 
addition, we manage to detect X-ray emission from the host galaxy (which 
is marked with an arrow in Fig.~2).

\pagebreak
\begin{wrapfigure}{i}[0pt]{62mm}
\centerline{\psfig{figure=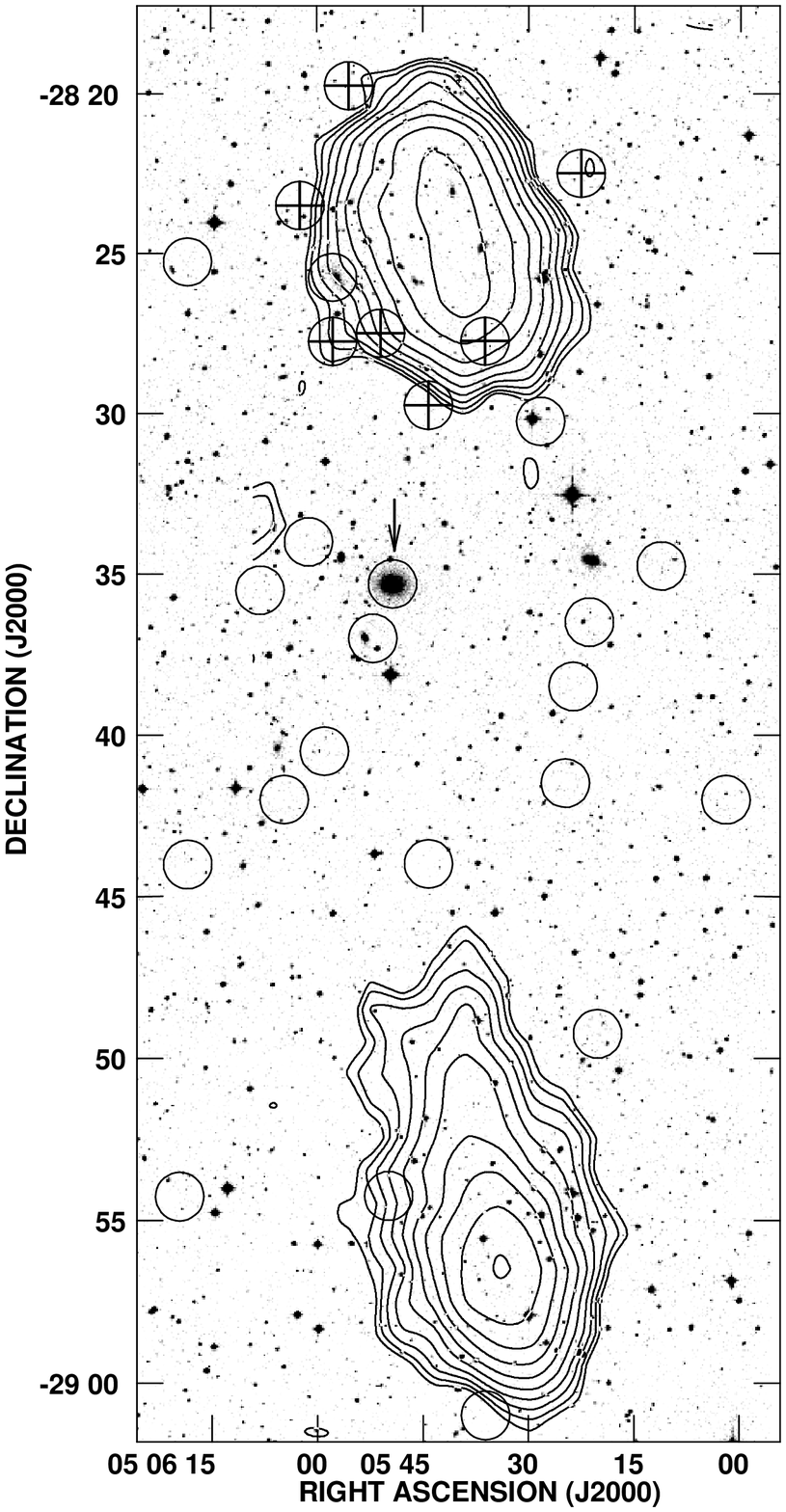,width=45truemm,angle=0,clip=}}
\captionb{2}{Positions of X-ray sources marked as circles (for details, see the text), superimposed on a 
gray-scale optical image and contours of the 6.3~cm radio emission.}
\end{wrapfigure}

Contrary to the symmetries of the radio structure mentioned above, the 
overall (projected) high-frequency morphology of B0503-286 is asymmetric 
in terms of the core-lobe arm-length ratio ($\sim$2) and the lobe 
morphology. The asymmetry of the radio giant could be caused by 
inhomogeneities in the ambient intergalactic medium, which is supported 
by the distribution of soft X-ray emission sources that show a significant 
excess in number density around the northern lobe. In addition, it seems that 
the brightest optically visible field galaxies tend to concentrate near the 
position of the host galaxy and in the vicinity of the northern lobe.

\vskip1mm
ACKNOWLEDGMENTS.\ \\MJ acknowledges the financial support from EAS.
\goodbreak

References\\
%\ref
Condon~J.~J., Cotton~W.~D., Greisen~E.~W., et al. 1998,  AJ, 115, 1693\\
%\ref
Fanaroff~B.~L., Riley~J.~M. 1974, MNRAS, 167, 31\\
%\ref
Kerp~J., Walter~F., Brinks~E. 2002, ApJ, 571, 809\\
%\ref
Klein~U., Mack~K.-H., Gregorini~L., Vigotti~M. 2003, A\&A, 406, 579\\
%\ref
Owen~F.~N., Ledlow~M.J. 1994, in {\it The First Stromlo Symposium: 
The Physics of Active Galaxies}, ASP Conf. Ser. 54, 
eds. G.~V.~Bicknell, M.~A.~Dopita, \& P.~J.~Quinn, p.~319\\
%\ref
Saripalli~L., Gopal-Krishna, Reich~W., K\"{u}hr~H. 1986, A\&A, 170, 20\\
%\ref
Subrahmanya~C.~R., Hunstead~R.~W. 1986, A\&A, 170, 27\\
%\ref
Trifalenkov~I.~A. 1994, AstL, 20, 215\\
%\ref
Veron-Cetty~M.~P., Veron~P. 2000, {\it A Catalogue of Quasars and Active 
Galactic Nuclei (9th Ed.)\\

\end{document}
\end